\documentclass[
showpacs,preprintnumbers,amsmath,amssymb]{revtex4}
\usepackage{graphicx}
\usepackage{dcolumn}
\usepackage{epsfig}
\usepackage{bm}
\def\beq{\begin{equation}}
\def\eeq{\end{equation}}
\def\be{\begin{equation}}
\def\ee{\end{equation}}
\def\bea{\begin{eqnarray}}
\def\eea{\end{eqnarray}}

\begin{document}

\title{Describing neutrino oscillations in  matter with
Magnus expansion
}

\author{A. N. Ioannisian$^{a,b}$,  A. Yu. Smirnov$^{c,d}$}
 \affiliation{
$^a$ Yerevan Physics Institute, Alikhanian Br.\ 2, 375036 Yerevan,
Armenia\\
$^b$ Institute for Theoretical Physics and Modeling, 375036 Yerevan,
Armenia\\
$^c$ International Centre for Theoretical Physics, Strada Costiera 11, 34014
Trieste, Italy\\
$^d$ Institute for Nuclear Research, Russian Academy of Sciences,
Moscow, Russia}


\begin{abstract}

We present new formalism for description of the  neutrino
oscillations in matter with varying density. The formalism is based on the
Magnus expansion and has a virtue that the unitarity of the S-matrix
is maintained in each order of perturbation theory.
We show that the Magnus expansion provides  better convergence of series:
the restoration of unitarity leads to
smaller deviations from the exact results especially  in the regions of
large transition probabilities.
Various expansions are obtained depending on a basis of neutrino states
and a way one  splits the Hamiltonian into the self-commuting and
non-commuting parts. In particular,  we develop the Magnus expansion for
the adiabatic perturbation theory which gives the best
approximation.  We apply the formalism to the neutrino
oscillations in matter of the Earth and show that
for the solar oscillation parameters the
second order Magnus adiabatic expansion has better than $1 \%$ accuracy
for all energies and trajectories.  For the atmospheric $\Delta m^2$ and
small 1-3 mixing the approximation works well
($< 3 \%$ accuracy  for $\sin^2 \theta_{13} = 0.01$)
outside the resonance region (2.7 - 8) GeV.

\end{abstract}



\maketitle

\section{Introduction}

Neutrino physics enters the era of precision measurements,
studies of the sub-leading oscillation effects and  searches for new physics beyond
the standard neutrino scenario. The neutrino
flavor conversions become a tool of exploration of other particles and objects such as
interiors of the Earth and stars. One of the key elements of these studies
is neutrino oscillations in matter
with varying density, and in particular, the oscillations inside the Earth.
The latter is relevant for the  solar, supernova and
atmospheric neutrinos, as well as for the cosmic and accelerator neutrinos.
In this connection it is important  to have  precise analytical
or semi-analytical expressions  for oscillation probabilities valid in wide
energy  ranges. These expressions allow us to  simplify
numerical computations  but also to gain a deeper insight into physics involved.
The results can be of special interest in view of discussions
of future experiments with the megaton-scale
fine structured underwater/underice detectors.

Several analytic and semi-analytic approaches to computing probabilities
in  matter with non-constant density have been
developed recently which use various perturbation theories
\cite{deHolanda:2004fd},
\cite{Ioannisian:2004jk},
\cite{Ioannisian:2004vv},
\cite{Akhmedov:2004rq},
\cite{Blennow:2004qd},
\cite{Akhmedov:2005yj},
\cite{LM},
\cite{Akhmedov:2006hb},
\cite{deAquino:2007sx},
\cite{Supanitsky:2007ks},
\cite{Liao:2007re}.
In the  previous publications
\cite{Ioannisian:2004jk}, \cite{Ioannisian:2004vv},  we have proposed a
formalism  which describes the neutrino oscillations in matter with low density.
It make use of  smallness of the matter potential $V$  in comparison with
the kinetic term: $V \ll   \Delta m^2/2E $, where $\Delta m^2$ is the mass
squared  difference and $E$ is the energy of neutrino.
Essentially,  the expansion parameter is given by the integral
along the trajectory
$$
I =  \int  dx \ V(x) \cos \phi (x),
$$
where $\phi (x)$ is the adiabatic phase.
The first approximation works very well at low energies
$E < 20$ MeV \cite{Ioannisian:2004jk}.
Validity of the results can be extended to higher energies if
the second order term,  $\sim I^2$, is taken into account \cite{Ioannisian:2004vv}.
It can be further improved  in certain energy ranges if
expansion is performed with respect to the deviation of the potential
from some average value.

The problem of this and some other similar approaches
is that the unitarity of oscillation amplitudes
is not guaranteed, and in fact,  is violated at high energies
\cite{Ioannisian:2004vv}.
This violation, in turn, can produce certain problems in numerical computations.
In this paper we propose the new type of perturbation theories which
maintain the unitarity explicitly in each order of expansion, and
therefore at any truncation of the series.
The approach is based on the Magnus expansion \cite{magnus}, \cite{burum}
which was previously used for description of the
nonadiabatic neutrino conversion in medium with monotonously varying density
\cite{D'Olivo:1990xs}, \cite{D'Olivo:1992mw}
\cite{AguilarArevalo:2003an}. Recently the first order Magnus expansion
has been applied  to the low energy neutrino oscillations in
matter of the Earth \cite{Supanitsky:2007ks}.
The formula for the regeneration factor
in the Earth has been obtained which generalizes our result in
\cite{Ioannisian:2004vv}.
In this paper we develop various perturbation theories using explicitly
two orders of the Magnus expansion. Since the Magnus expansion is an expansion
in power of commutators, it is the second order that provides non-trivial new results.
As a part of the present study we reproduce
the formula from \cite{Supanitsky:2007ks}.

Essentially, the restoration of unitarity in the Magnus expansion is achieved by
an effective re-summation of certain contributions to oscillation amplitudes.
This leads to higher accuracy of the semi-analytic results
and  allows us to further extend the range of applications of the approach.
Furthermore, it gives better understanding of the previously obtained results
and their limits of validity.\\

We illustrate an accuracy of the approximations
computing the transition probabilities for neutrinos
crossing the core of the Earth. We find that for the
solar oscillation parameters the second order
Magnus adiabatic expansion has better than $1 \%$ accuracy
for all energies and all trajectories.  For the atmospheric $\Delta m^2$ and
small 1-3 mixing the approximation works very well ($< 3 \%$ accuracy for $\sin^2
\theta_{13} = 0.01$) below  2.7 GeV and above  8 GeV for $\sin^2 \theta_{13} = 0.01$.
In the region,  (2.7 - 8) GeV,  where the MSW resonances
in the core and in the mantle as well as the parametric
resonances take place, a special consideration is required.

The paper is organized as follows. In sec. 2 we present the
formalism of Magnus expansion and
obtain  general expressions for the S-matrix.
We calculate the oscillation probabilities
using various perturbation approaches
based on the Magnus expansion in sec. 3. In particular, we develop the
perturbation theory in $I$ and the adiabatic perturbation theory.
We compare the results of different semi-analytic approaches in sec. 4.
Conclusions follow in sec. 5.

\section{Magnus expansion}

\subsection{S-matrix and Magnus expansion}

In what follows we will mainly study the case of $2 \nu-$mixing
$(\nu_e, \nu_\alpha)$, where $\nu_\alpha$ is, in general,
   some combination of $\nu_\mu$ and $\nu_\tau$.
In a number practical cases the two neutrino results can be immediately
embedded in the complete $3\nu$ mixing scheme.

The evolution matrix of neutrinos in matter, $ S(x, x_0)$, obeys
the first order (operator) differential equation, \be \label{schr}
i {d \ S (x, x_0) \over d \ x} = H(x) S(x, x_0) \ , \ee where the
Hamiltonian $H(x)$ is given in the flavor basis by
 \be
H = \frac{M M^{\dagger}}{2E} +{\hat{V} } = \frac{1}{2E} U (\theta)
M_\Delta^2 U(\theta)^{\dagger} + {\hat{V}}.
 \ee
Here ${\hat{V}} \equiv diag(V, 0)$ is the matrix of potentials,
\be U (\theta) \equiv \left( \hspace{-0.2cm}
\begin{tabular}{rc}
$\cos \theta$ & $\sin \theta$ \\
$-\sin \theta$ & $\cos \theta$
\end{tabular}
\right)
\label{v-mix}
\ee
is the mixing matrix, and
$M_\Delta^2 \equiv diag (0, \Delta m^2)$
is the diagonal matrix of mass squared differences.

Formally,  the solution of the equation  (\ref{schr}) can be written
as the chronological product
\be
  S (x_f, x_0) =  T e^{-i\int_{x_0}^{x_f}H(x) \ dx }\equiv
  \lim_{n \to \infty}
   e^{-i H(x_n) \Delta x} \cdot
   e^{-i H(x_{n-1}) \Delta x}\cdots
   e^{-i H(x_1) \Delta x} \ ,
\label{s-matr}
\ee
$$
\Delta x = {x_f-x_0 \over n} \ .
$$
In our previous papers, \cite{Ioannisian:2004jk}, \cite{Ioannisian:2004vv},
we performed expansion of each  exponential
factor in eq. (\ref{s-matr}) and then took limit $n  \rightarrow \infty$.
Such a procedure does not guarantee the unitarity once the series is truncated and
finite number of terms of  the expansion is taken.

In this paper we will use expansions of powers of
exponents and sum up contributions in the power
without expansion of exponents themselves. Consequently, the form,
$S =  e^{-i C}$, of the  $S$-matrix,
and therefore, the unitarity are maintained since $C$ is a hermitian matrix.
The  Magnus expansion~\cite{magnus} has the following form
\be
S =   e^{-i C [H]} \equiv  e^{-i (C_1 + C_2 + C_3 + ...)},
\label{Crepres1}
\ee
where functional $C[H]$ is a series in powers
of commutators of the Hamiltonians taken in different points of
neutrino trajectory.  The term  $C_k [H] $ contains commutators
of order $k-1$:
\begin{eqnarray}
\label{c111}
C_1&=&\int_{x_0}^{x_f} \! \! \! \! \! dx \ H(x)~ ,
\\
C_2&=& {-i \over 2}\int_{x_0}^{x_f}\! \! \! \! \! dx\int_{x_0}^x
\! \! \! \! \! dy  \
[H(x), \ H(y)]~,
\label{c222}\\
C_3&=& {(-i)^2 \over 6} \! \!  \int_{x_0}^{x_f}\! \! \! \! \! dx
\int_{x_0}^x \!\! \! \! \! dy  \int_{x_0}^y \! \! \! \! \! dz \left(
[H(x), [H(y),  H(z)]] + [[H(x),  H(y)],  H(z)] \right).
\label{c333}
\end{eqnarray}
The details of derivation of the functionals $C_k[H]$ are given in the Appendix.
The representation of the $S$ matrix in eq.(\ref{Crepres1})
with  $C_i$ given in (\ref{c111}) (\ref{c222}), (\ref{c333})
is the  main tool which we will use for different applications.

The Magnus expansion is an  integral version of
the Baker-Campbell-Hausdorff (BCH) equality.
Recall that according to the BCH-equation,
the summation of  powers in exponents leads to
$$
 e^a \cdot e^b = e^{a+b+{1\over 2} [a,b]+ {1 \over 12} \left[(a - b) [a,b] \right]
+ \dots},
$$
that is, to appearance of commutator of the operators.
In fact,  in  matter with varying density the
Hamiltonians taken in different spatial points do not commute
$$
[H(x_i) , H(x_j)] \not = 0.
$$
The calculation of the S-matrix (\ref{s-matr}) requires
an extension of the BCH-equality to
a product of many exponential factors, and eventually, a transition
to the continuous limit.

\subsection{Properties of Magnus expansion.}

Let us consider general properties of the Magnus expansion
given in eqs.(\ref{Crepres1}, \ref{c111}, \ref{c222}, \ref{c333}). \\

1). If $H(x) =$ constant,  then  $C_i = 0$ for $i>1$, and therefore in the uniform
medium
the $S-$ matrix is given by
$$
S = e^{-i\int_{x_0}^{x_f}  dx \ H(x)} = e^{-i H ~(x_f - x_0)}.
$$
This reproduces immediately the standard oscillation results.
All corrections (due to the non-constant Hamiltonian) are given by
the commutators.
Essentially,  the Magnus expansion is the expansion in the number of commutators. \\

2). The terms of the Magnus expansion
(\ref{c111}, \ref{c222}, \ref{c333})
contain factorials in denominator, therefore
a  convergence of the series is better than a convergence of the usual expansion
(see eq. (\ref{t-prod}) in the Appendix).
The Magnus series has good convergence even if $H$ is not small. \\

3). The commutators themselves may contain
an additional smallness. The weaker dependence of $H$ on distance the smaller the
commutators. So, in a sense,   we deal here with a kind of adiabatic expansion.\\

4). If $H(x)$ is a symmetric function with respect to the
middle point of a neutrino trajectory,
$$
{\bar x} = {x_f+x_0 \over 2 },
$$
that is,
\be
H(x) = H(2 \bar{x} - x),
\label{symmtra}
\ee
one can show that $C_{2n}=0$ ($n=1,2...$) \cite{burum}, and  only the odd
terms in the expansion are non-zero.
Let us prove that $C_2 = 0$ (general proof is given in \cite{burum}).
According to eq. (\ref{c222}) the integration region $(y = x_0 \div x,~~ x = x_0
\div
x_f)$
is symmetric with respect to the diagonal line $y = 2 \bar{x} - x$,
that is, symmetric under reflection:
\be
(x, ~y) \rightarrow (2 \bar{x} - y, ~ 2 \bar{x} - x)
\label{refl}
\ee
($x > y$).  Taking into account the symmetry of Hamiltonian  (\ref{symmtra})
it is easy to show that under the reflection (\ref{refl}) the
commutator $[H(x), H(y)]$ changes the sign. Therefore
the integration of this commutator gives zero.

\subsection{Magnus expansion in the ``interaction'' representation}

Let us split the total Hamiltonian into two parts
\be
H(x) = H_0(x)+\Upsilon(x)
\label{hsplit}
\ee
in such a way that $H_0(x)$ is self-commuting along a trajectory. That is,
for any two points of the trajectory $x_i$, $x_j$:
$[H_0(x_i), \ H_0(x_j)]=0$.  The rest of the Hamiltonian, $\Upsilon(x)$,
is not self-commuting, in general, and if small can be treated as a perturbation.
In this case it is convenient to solve the problem in the basis
of new states, $\psi_I$,  related to the  initial basis by
\be
\psi =  U_I(x) \psi_I =  e^{-i \int_{x_0}^{x} dt H_0 (t)} \psi_I .
\label{intrep}
\ee
Inserting this relation  into the evolution equation we
find that $\psi_I$, and the corresponding $S-$matrix,  satisfy
the evolution equation with the Hamiltonian $H_I \equiv \Upsilon_I$, where
\be
\Upsilon_I (x, x_0) =
U_I^\dagger \Upsilon(x) U_I(x) =
e^{i\int_{x_0}^x H_0(t) \ dt} \Upsilon(x)
e^{-i\int_{x_0}^x H_0(t) dt}.
\ee
The  transformation to new basis (\ref{intrep}) is equivalent to
transition to a ``interaction representation'' if
$H_0$ is interpreted as the Hamiltonian of free propagation.
$\Upsilon_I$ can be considered as an operator
in the interaction representation.

The evolution matrix in the interaction representation is given by
\be
S_I(x_f, x_0) = e^{-iC[\Upsilon_I (x, x_0)]},
\ee
that is, in the formulas (\ref{c111}, \ref{c222}, \ref{c333}) one should substitute
$H(x) \rightarrow  \Upsilon_I (x, x_0)$.
Then, according to eq. (\ref{intrep}) the $S-$ matrix in the original basis equals
\be
S(x_f, x_0) =   U_I(x_f) S_I(x_f, x_0) U_I(x_0)^{\dagger} ,
\ee
or explicitly,
\be
S(x_f, x_0) =   e^{-i \int_{x_0}^{x_f} dt H_0 (t)}  e^{-iC[\Upsilon_I (x, x_0)]}.
\label{orig-bas}
\ee
(The exponent on the RH side of this equality disappears because of the
integration limits.)
If  $\Upsilon(x) \ll H_0(x)$, so that it can be considered as a small perturbation,
a convergence of the series will be fast. \\

The Hamiltonian is self-commuting if its
dependence on distance can be factorized:
\be
H_0(x) = f(x)\cdot M,
\label{factor-m}
\ee
here $f(x)$ is an arbitrary function of $x$
and $M$ is an arbitrary constant matrix.
Specific realizations of (\ref{factor-m}) include constant ($x$-independent)
Hamiltonians as well as  the diagonal Hamiltonians  $H_0(x) = diag [f_1 (x), f_2
(x)]$. In the latter case subtracting a
matrix proportional to the unit matrix:  $0.5 (f_1 + f_2) diag (1, 1) $,
one can reduce the Hamiltonian to  the form (\ref{factor-m}).

In the case of small mixing (which can be achieved selecting certain  basis
of neutrino states) we can split the Hamiltonian as
$$
H (x) = H^{diag}(x) + H^{off-diag}(x)
$$
and identify $H^{off-diag}(x)$ with $\Upsilon$.

\subsection{Evolution in  symmetric potential  }

Let us consider a symmetric density profile so that
the Hamiltonian satisfies the equality (\ref{symmtra}).
In this case it is  convenient to perform the integration in $C_i$ from
the middle point  of neutrino trajectory, $\bar{x}$,
and to choose the evolution basis $\psi_I$,
such that $\psi = \bar{U}_I \psi_I$ with
\be
\bar U_I(x) =  e^{-i \int_{\bar x}^{x} dt H_0 (t)} .
\label{av-rep}
\ee
Essentially here we have substituted $x_0$ by $\bar{x}$.
Now (similarly to the consideration in the previous section)
the evolution matrix can be written as
\be
S_I (x_f, x_0) = e^{-iC[\Upsilon_I (x, \bar{x})]},
\ee
where
\be
\Upsilon_I (x, \bar{x}) =
e^{i\int_{\bar{x}}^x H_0(t) \ dt} \Upsilon(x)
e^{-i\int_{\bar{x}}^x H_0(t) dt}.
\ee
Then, the evolution matrix in the original basis equals
 \be
S(x, x_0) =   \bar{U}_I(x) S_I (x, x_0) \bar{U}_I(x_0)^{\dagger} ,
 \ee
or explicitly,  for an  evolution from $x_0$ to  $x_f$  we obtain
\be
S(x_f, x_0) =   e^{-i \int_{\bar{x}}^{x_f} dt H_0 (t)}
e^{-iC[\Upsilon_I (x, \bar{x})]} e^{-i \int^{\bar{x}}_{x_0} dt H_0 (t)} .
\label{s-origin}
\ee

Notice that in contrast to $\Upsilon(x)$ the operator
$\Upsilon_I(x)$ has no definite symmetry with respect to the
middle of a trajectory even for a constant density  profile. Therefore
the even coefficients, $C_{2k}$,  are non-zero:
\begin{eqnarray}
\bar C_1 \equiv C_1[\Upsilon _I (x, \bar{x})]  \!\!\!\! & = & \!\!\!\!  \int_{x_0}^{x_f} \!\!\!\! dx
\Upsilon_I(x),
\nonumber\\
\bar{C}_2 \equiv C_2[\Upsilon _I (x, \bar{x})] \!\! & = &  -i {1 \over 2} \int_{x_0}^{x_f} \! \! \!
\! \! dx \!\! ~ \int_{x_0}^x \! \! \! \! \! dy
\left[ \Upsilon_I (x),~~\Upsilon_I (y) \right],
\label{cc1122}
\end{eqnarray}
{\it etc.}. Here ``bar'' indicates
that $\bar{C}_i$ have been calculated in the interaction
representation with the $\bar{U}_I$-matrix
integrated from the middle point of  trajectory.

Let us introduce the variable
\be
r \equiv x - \bar{x} =  x -{x_f+x_0 \over 2}
\ee
which is the distance from the middle of trajectory.
 Then
\begin{eqnarray}
\bar{C}_1   & = & \int_{-L}^{L} \!\!\  dr \Upsilon_I(r),
\label{c1-bar}\\
\bar{C}_2  & =&\!\! - {i \over 2} \int_{-L}^{L} \!\! dr \! \! \!
\int_{-L}^{r} \!\!\  dp \left[ \Upsilon_I (r),~~\Upsilon_I (p)
\right]. \label{c2-bar}
\end{eqnarray}
Here
 \be
L \equiv \frac{x_f - x_0}{2}
 \ee
and
\be
\Upsilon_I (r) =
e^{i\int_{0}^r H_0(t) dt} \Upsilon (r)
e^{-i\int_{0}^r H_0(t) dt}.
\label{ups-r}
\ee
Notice that the expressions (\ref{c1-bar}, \ref{c2-bar}, \ref{ups-r})  are valid
for any density profile
and we have not used  yet any symmetry of the Hamiltonian.\\

Let us  now assume that $V(x)$, and consequently,
the Hamiltonian, are symmetric functions with respect to
the middle point of a trajectory, $r = 0$,
(as for neutrinos  crossing  the Earth). In this case $H_0$ and $\Upsilon$ are
the even functions of $r$:
\be
H_0 (- r) = H_0 (r), ~~~ \Upsilon (-r) = \Upsilon (r).
\ee
Denoting
\be
\Phi_0 \equiv  \int_{0}^r H_0(t) dt
\ee
we have
\be
\Phi_0 (- r)  =  - \Phi_0 (r)
\label{phase-m}
\ee
provided that $H_0$  is real.
Let us show that in this case $\bar{C}_1$ and  $ \bar{C}_2$ are
the real symmetric matrices.
The proof is straightforward in the case of real $\Upsilon$.
The function  $\Upsilon_I (r)$ is not symmetric  with respect
to $r = 0$. Indeed,  rewriting (\ref{ups-r})
as
\be
\Upsilon_I (r) =  e^{i\Phi_ 0(r)} \Upsilon (r) e^{-i\Phi_0 (r)},
\label{ups-r1}
\ee
one can see immediately that under $r \rightarrow - r$
\be
\Upsilon_I (- r) =   \Upsilon_I (r)^*.
\label{Ups-sym}
\ee
Using this relation and the definition (\ref{c1-bar}) we obtain
$$
\bar{C}_1^* = \int_{-L}^{L} dr \Upsilon_I(r)^* = \int_{-L}^{L} dr
\Upsilon_I(- r) = \int_{-L}^{L} dr \Upsilon_I(r) = \bar{C}_1 ,
$$
where in the last equality we made a substitution $r \rightarrow - r$.
Furthermore, since  $\bar{C}_1$ is Hermitian,   $\bar{C}_1 = \bar{C}_1^{\dagger}$,
we obtain that  $\bar{C}_1 = \bar{C}_1^T$, {\it i.e.}, the matrix is symmetric.

Similarly we can show that $\bar{C}_2^* = \bar{C}_2$. Here in addition to the
property (\ref{Ups-sym}) and the change of the signs of variables,
we use that
$$
\int_{-L}^{L} dr \int^{L}_{r}   dp \left[ \Upsilon_I
(r),~~\Upsilon_I (p) \right] = - \int_{-L}^{L} dr \int_{-L}^{r} dp
\left[ \Upsilon_I (r),~~\Upsilon_I (p) \right].
$$
Again, since  $\bar{C}_2$ is Hermitian,
the matrix $\bar{C}_2$ should be symmetric.

Performing integration in the expressions for $\bar{C}_i$
(\ref{c1-bar}, \ref{c2-bar}) from the middle point of a trajectory we obtain
\begin{eqnarray}
\bar{C}_1 & = & 2 \int_{0}^{L}  dr {\rm Re} \Upsilon_I(r),
\nonumber \\
\bar{C}_2 & = & 2 \int_{0}^{L}  dr \int_{0}^{r} \!\!\  dp \left[
{\rm Im} \Upsilon_I (r),~~ {\rm Re} \Upsilon_I (p) \right]
\label{c12-mid}
\end{eqnarray}
from which we immediately conclude that $\bar{C}_i$  are real.

As we will see in sect. III.C,  in the adiabatic perturbation theory
$\Upsilon$ is purely imaginary matrix. Moreover, since $\Upsilon \propto dV/dx$,
for a symmetric potential we have the antisymmetric  $\Upsilon$. So,
\be
\Upsilon (r)^* = - \Upsilon (r), ~~~\Upsilon (- r) = - \Upsilon (r),
\label{upseq}
\ee
and therefore $\Upsilon (- r) =  \Upsilon (r)^*$.
Using the equalities  (\ref{upseq}) one can show that
in this case  $\Upsilon_I (r)$
also satisfies the equality
(\ref{Ups-sym}), and consequently the matrices $\bar{C}_i$
can be calculated as in  eq. (\ref{c12-mid}).

For a symmetric potential using the property (\ref{phase-m}) we can
write the S-matrix in the original basis (\ref{s-origin}) as \be
S(x_f, x_0) =   e^{-i \Phi_0 (L)} e^{-iC[\Upsilon_I (x, \bar{x})]}
e^{-i \Phi_0 (L)}. \label{s-origins} \ee

\section{Oscillation probabilities}

In applications of the Magnus expansion,
adjusting the formalism to a specific physical situation
we can select
\begin{itemize}

\item
propagation basis,  that is, the basis of neutrino states
 in which we consider evolution;

\item
split of the Hamiltonian into
 self-commuting and non-commuting parts;

\item
perturbation terms.

\end{itemize}

In what follows we will consider a symmetric density profile keeping in mind
applications to the neutrino propagation inside the Earth.

\subsection{Low energy and low density limit}

In the low energy or/and low density case it is convenient to consider
the neutrino evolution in the mass eigenstates basis, $\nu_{mass} = (\nu_1, \nu_2)$.
In this basis the Hamiltonian  can be written as
\be
 H(x) =\left(
\begin{tabular}{cc}
0 & 0 \\
0 & $\Delta m^2/2E$
\end{tabular}
\right)+ U^\dagger \left(
\begin{tabular}{cc}
$V(x)$ & 0 \\
0 & 0
\end{tabular}
\right) U,
\label{ham10}
\ee
where $U$ is the vacuum mixing matrix (\ref{v-mix}).
We split the Hamiltonian, according to (\ref{hsplit}),  in the following way.
The self-commuting part can be chosen  as
\begin{eqnarray}
H_0(x) =\left(
\begin{tabular}{cc}
0 & 0 \\
0 & $\Delta^m (x)$
\end{tabular}
\right),
\label{h0}
\end{eqnarray}
where  $\Delta^m(x)$ is the difference of the instantaneous
eigenvalues of the Hamiltonian (\ref{ham10}):
\begin{equation}
\Delta^m(x)  \equiv {\Delta m^2 \over 2E}
           \sqrt{\left(\cos 2\theta- \frac{2EV(x)}{\Delta m^2}\right)^2 +\sin^2 2 \theta}~ .
\label{delta}
\end{equation}
Then, according to (\ref{ham10}), the perturbation part equals
\begin{eqnarray}
\hspace{-0.4cm} \Upsilon(x) =
A(x) \left(
\begin{tabular}{cc}
0 & 1 \\
1 & 0
\end{tabular}
\right) +
B(x) \left(
\begin{tabular}{cc}
1 & 0 \\
0 & -1
\end{tabular}
\right),
\label{ups}
\end{eqnarray}
where
\bea
A (x) & \equiv & \frac{1}{2} \sin 2\theta ~V(x),
\nonumber\\
B(x) & \equiv & {1 \over 2} \left[\Delta^m (x)- \frac{\Delta m^2}{2E} + V(x)
\cos 2\theta \right].
\label{aandb}
\eea
For a weak potential $V$: $V \ll \Delta m^2/ 2E$, we have
\be
B(x) = \frac{1}{4}(V \sin 2 \theta)^2 \frac{2E}{\Delta m^2} + O(V^3)
\approx A^2 (x) \frac{2E}{\Delta m^2}.
\label{Bappr}
\ee
According to (\ref{h0}) the matrix of transition
to the interaction representation equals
\be
\bar{U}_I (x) =
\left(\begin{tabular}{cc}
1 & 0 \\
 0 & $e^{-i \phi (x)}$
\end{tabular}
\right),
\label{baru}
\ee
where
\be
\phi(x) \equiv  \int_0^x \Delta^m(r) \ dr
\ee
is the adiabatic phase (here the integration runs from
the middle point of a trajectory).
Then the Hamiltonian in the interaction representation, $\Upsilon_I(x) =
\bar{U}^{\dagger}(x) \Upsilon (x) \bar{U} (x)$  can be written as
\begin{eqnarray}
\hspace{-0.4cm}
\Upsilon_I (x) =
A(x) \left(
\begin{tabular}{cc}
0 & $e^{-i \phi (x)}$ \\
$e^{i \phi (x)}$ & 0
\end{tabular}
\right) +
B(x) \left(
\begin{tabular}{cc}
1 & 0 \\
0 & -1
\end{tabular}
\right).
\label{upsI}
\end{eqnarray}
Using this expression and  eqs. (\ref{c12-mid}) we obtain
\begin{eqnarray}
\bar C_1 + \bar C_2  = Z(L) \left(
\begin{tabular}{cc}
0 & 1 \\
1 & 0
\end{tabular}
\right) + Y(L) \left(
\begin{tabular}{cc}
1 & 0 \\
0 & -1
\end{tabular}
\right),
\label{upscc}
\end{eqnarray}
with
 \bea
Z(L) & \equiv & 2\int_0^{L} dr A(r)\cos \phi (r)  + 4 \int_0^{L}
dr \int_0^r dp~ A(r) B(p) \sin\phi(r) ,
\nonumber\\
Y(L) & \equiv & 2\int_0^{L} dr B(r) - 4 \int_0^{L} dr \int_0^r dp~
A(r) A(p) \sin\phi(r) \cos \phi(p). \label{yyzz}
 \eea

Let us estimate these quantities with accuracy $\sim V^2$. Since
$A \sim V$ and $B \sim V^2$, the last term in $Z$, being of the
order $V^3$,  can be neglected. For the function $Y (x)$
performing integration by parts in the second integral
we have
 \bea
Y (L)  & = & 2\int_0^{L} dr \left[B(r)  - 2  {A(r)^2 \over \Delta
^m(r)} \right] + 4 \int_0^{L} dr \int_0^r dp {d \over
dr}\left[{A(r) \over \Delta^m(r)}\right] {d \over dp}\left[{A(p)
\over \Delta^m(p)}\right] \sin\phi(p) \cos \phi(r)
\nonumber\\
&=& \sin^2 2 \theta \int_0^{L} dr \int_0^r dp {d \over
dr}\left[{V(r) \over \Delta^m(r)}\right] {d \over dp}\left[{V(p)
\over \Delta^m(p)}\right] \sin\phi(p) \cos \phi(r) +   O(V^3),
\label{yyy}
 \eea
where in the last equality  we used expression (\ref{Bappr}).

Neglecting $Y(x)$ we find
\begin{equation}
\bar C \simeq  \bar C_1 + \bar C_2 \approx I_V \left(
\begin{tabular}{cc}
0 & 1 \\
1 & 0
\end{tabular}
\right),
\label{barc-11}
\end{equation}
and
 \be
I_V  \equiv \sin 2 \theta \int_{0}^{L} \!\!\!\! dr \ V(r) \cos
\phi (r).
 \ee
Using eqs. (\ref{s-origins}), (\ref{barc-11}) and (\ref{baru})  we obtain
the $S-$matrix (\ref{orig-bas}) in the mass-eigenstates basis
\begin{eqnarray}
S &=& \left(
\begin{tabular}{cc}
1 & 0 \\
 0 & $e^{-i \phi}$
\end{tabular}
\right)
\left(
\begin{tabular}{cc}
$\cos I_V$ & $ -i \sin I_V$ \\
 $-i \sin I_V$ & $\cos I_V$
\end{tabular}
\right) \left(
\begin{tabular}{cc}
1 & 0 \\
 0 & $e^{-i \phi}$
\end{tabular}
\right).
\label{smat-mass}
\end{eqnarray}
Here $\phi$ is the half of the oscillation phase:
 \be
 \phi \equiv \phi_{\bar x \to x_f} =\phi_{x_0 \to \bar x} = \phi(L).
\label{phasesh}
 \ee
Notice that both the matrix  that originates from the
self-commuting part and the perturbation, $I_V$, depend on the
same adiabatic phase.

For the transition between the mass states we have immediately from
(\ref{smat-mass}):
\be
P_{\nu_2 \to \nu_1} =  |S_{21}|^2 =  \sin^2 I_V.
\ee
The $S-$matrix for the mass-to-flavor transitions equals
$$
S_{mass - flavor} = U (\theta) \cdot S,
$$
and the $\nu_i \to \nu_\alpha$ probability is
\be
P_{\nu_i \to \nu_\alpha} = |(U \cdot S)_{\alpha i}|^2 .
\label{mtof}
\ee
\, From (\ref{mtof}), (\ref{smat-mass}) and
(\ref{v-mix}) we obtain
\begin{equation}
P_{\nu_2 \to \nu_e} = \sin^2\theta + \frac{1}{2} \sin 2 \theta ~ \sin 2 I_V
~\sin \phi + \cos 2\theta ~ \sin^2 I_V ,
\label{2eresum}
\end{equation}
where the first term is simply projection squared of $\nu_2$ state onto $\nu_e$.
Eq. (\ref{2eresum}) reproduces the formula given in \cite{Supanitsky:2007ks}.
If $|I_V | <<1$, we find making expansion in powers of $I_V$
\begin{eqnarray}
P_{\nu_2 \to \nu_e}
&=& \sin^2 \theta + I_V \sin 2 \theta \sin \phi + I_V^2 \cos 2 \theta
\label{old}
\end{eqnarray}
which exactly coincides with
our result in \cite{Ioannisian:2004vv} (see eq. (15)).
In a sense, the result (\ref{2eresum})
corresponds to a re-summation of certain
contributions to the probability.  It is the
substitution $I_V \rightarrow \sin I_V$ that restores the unitarity. Notice that
$I_V = \sin 2\theta~I$, where
$I$ is the integral used as the expansion parameter in \cite{Ioannisian:2004vv}. According to
the  present result (\ref{2eresum})  the expansion parameter includes also $\sin 2\theta$ which
makes convergence even better in the case of small vacuum mixing. Our present consideration
explains also the reason why the second order effect in ref.  \cite{Ioannisian:2004vv} depends
on the same integral $I$.

The $S-$matrix for transitions between the flavor states equals
$$
S_{flavor - flavor} = U \cdot S \cdot U^{\dagger}.
$$
In particular, for the $\nu_e \to \nu_\alpha$  channel we obtain
\bea P_{\nu_e \to \nu_\alpha} & = & \cos^2 I_V ~\sin^2 2\theta ~
\sin^2 \phi + {1 \over 2 } \sin 2I_V \ \sin 4 \theta \ \sin\phi \
+
 \sin^2 I_V \ \cos^2 2\theta
\nonumber\\
~~ & = & (\cos I_V  ~\sin 2\theta ~ \sin \phi + \sin I_V \ \cos
2\theta)^2. \eea In the limit $V \rightarrow 0$, we have  $I_V
\rightarrow 0$ and the first term reproduces the standard vacuum
oscillation probability. For small $I_V$ the following form of the
probability can be useful:
\be
P_{\nu_e \to \nu_\alpha}  =  \sin^2
2\theta ~ \sin^2 \phi + {1 \over 2 } \sin 2I_V \ \sin 4 \theta \
\sin\phi \ +
 \sin^2 I_V ( \cos^2 2\theta - \sin^2 2\theta ~ \sin^2 \phi).
\ee

The result in the second order of the Magnus expansion can be obtained
keeping term proportional to $Y(x)$ in $\bar{C}$ (\ref{upscc}).
Straightforward calculations give
\be
S  =
\left(
\begin{array}{cc}
\cos X  - i \frac{Y}{X} \sin X  & -i e^{-i\phi} \frac{Z}{X}  \sin X  \\
 -i e^{-i\phi}\frac{Z}{X}  \sin X   &
  e^{-2i\phi} \left[ \cos X + i \frac{Y}{X} \sin X \right]
\end{array}
\right),
\label{smatxyz}
\ee
where $X \equiv \sqrt{Z^{2} + Y^2}$.
Apparently, the result (\ref{smat-mass}) follows from this
expression in the limit $Y \rightarrow 0$, $Z \rightarrow I_V$.

\subsection{Perturbation around average potential $V_0$}

Let us consider the same situation as in the previous section but perform
the expansion with respect to an average potential  $V_0$.
This means that we use
the basis of neutrino eigenstates in matter with constant potential
$V_0$, as the propagation basis.
These eigenstates are related to the flavor states
by the mixing matrix in matter
\be
\nu_f = U(\theta^m_0) \nu^m_0,
\ee
where $U$ is defined in (\ref{v-mix}) and
$\theta^m_0 = \theta^m (V_0)$ is
the mixing angle in matter with the potential $V_0$,
the angle $\theta^m (V)$ is given by
\be
\sin 2 \theta^m (V) =
{\sin 2 \theta \over \sqrt{(\cos 2 \theta - 2E V /\Delta m^2)^2+\sin^2 2 \theta}}.
\label{mangle}
\ee
In the $\nu^m_0$- basis the Hamiltonian equals
\be
 H(x) =\left(
\begin{tabular}{cc}
0 & 0 \\
0 & $\Delta_0^m$
\end{tabular}
\right)+ U^\dagger (\theta^m_0) \left(
\begin{tabular}{cc}
$\Delta V(x)$ & 0 \\
0 & 0
\end{tabular}
\right) U(\theta^m_0),
\label{ham1}
\ee
where $\Delta_0^m$ is the difference of the eigenvalues in matter with
the potential $V_0$, and
$$
\Delta V(x) \equiv V(x) - V_0.
$$

We split the Hamiltonian into the self-commuting part and the perturbation using
the same $H_0$ as in the previous case (\ref{h0}).
Then the perturbation equals
\begin{eqnarray}
\hspace{-0.4cm} \Upsilon(x) = \frac{1}{2} \sin 2\theta ~ \Delta V(x) \left(
\begin{tabular}{cc}
0 & 1 \\
1 & 0
\end{tabular}
\right) + {1 \over 2} \left[\Delta^m  - \Delta_0^m + \Delta V(x)
\cos 2\theta_0^m \right] \left(
\begin{tabular}{cc}
1 & 0 \\
0 & -1
\end{tabular}
\right).
\label{ups1}
\end{eqnarray}
Consequently, for the matrix $\bar{C}$ we obtain the same expression as in
Eq. (\ref{barc-11}) with substitution $I_V \rightarrow I_V^\prime$, where
\be
    I_V^\prime =\sin 2 \theta^m_0 \int_{0}^{L}
    \Delta V(x) \ \cos \phi (x) \ dx .
\ee
In turn,   $I_V^\prime$ differs from  $I_V$ by the
substitutions $V \rightarrow \Delta V$ and
$\theta \rightarrow \theta^m_0$.

The $S-$matrix in the $\nu^m_0-$basis equals
$$
S_0^m =
\left(
\begin{tabular}{cc}
$\cos I_V^\prime$ & $ -i e^{-i \phi}  \sin I_V^\prime$ \\
 $-i e^{-i \phi} \sin I_V^\prime$ & $ e^{-2 i\phi}\cos I_V^\prime$
\end{tabular}
  \right),
$$
and the phase $\phi$ is defined in (\ref{phasesh}).

Since $\nu = U^{\dagger}(\theta) U(\theta_0^m)\nu^m_0 =
U(\theta_0^m - \theta)\nu^m_0$,
the $S-$matrix of  the mass-to-flavor transitions equals
$$
S_{mass - flavor} =
    U(\theta^m_0) \cdot S_0^m  \cdot
    U(\theta^m_0-\theta)^\dagger .
$$
Then the $\nu_2 \to \nu_e$ probability
is
\bea
P_{\nu_2 \to \nu_e} =
\cos^2 I_V^\prime [ \sin^2 \theta
 + \sin 2\theta^m_0 \sin 2(\theta^m_0-\theta) \sin^2 \phi] +
 \nonumber \\
 +  {1 \over 2 } \sin 2 I_V^\prime  \sin 2 (2\theta^m_0-\theta) \sin \phi
+ \sin^2 I_V^\prime  \cos^2 (2 \theta^m_0- \theta).
\label{mf-shift}
\eea
Apparently this expression is reduced to the one in eq. (\ref{2eresum}),
if $\theta^m_0 = \theta$.

For the mass-to-mass transitions the $S-$matrix equals
$$
S_{mass - mass} =
    U(\theta^m_0 -\theta) \cdot S_0^m  \cdot
    U(\theta^m_0-\theta)^\dagger ,
$$
and therefore the  probabilities are given by the same expressions as  for the
flavor-to-flavor transitions in the previous section with the
substitutions $\theta \rightarrow (\theta^m_0 -\theta)$  and $I_V
\rightarrow I_V^{\prime}$:
\bea
    P_{\nu_2 \to \nu_1} &=&
 \cos^2 I_V^\prime  \sin^2 2 (\theta^m_0-\theta)  \sin^2 \phi +
{1 \over 2 } \sin 2 I_V^\prime  \sin 4 (\theta^m_0-\theta)  \sin \phi
 + \sin^2 I_V^\prime  \cos^2 2( \theta^m_0-\theta)
\nonumber\\
& = & \left[ \cos I_V^\prime \ \sin 2 (\theta^m_0- \theta) \ \sin \phi +
   \sin I_V^\prime \ \cos 2( \theta^m_0-\theta) \right]^2 .
\label{21pprime}
\eea
For the flavor-to-flavor transition we have
$$
S_{flavor - flavor} =
    U(\theta^m_0) \cdot S_0^m  \cdot
    U(\theta^m_0)^\dagger .
$$
Consequently, the probability follows immediately from (\ref{21pprime})
substituting $(\theta^m_0 - \theta) \rightarrow \theta^m_0$:
\bea
 P_{\nu_e \to \nu_\alpha} & = &
 \cos^2 I_V^\prime \ \sin^2 2\theta^m_0 \ \sin^2 \phi
+   {1 \over 2 } \sin 2 I_V^\prime \ \sin 4\theta^m_0 \ \sin \phi  +
\sin^2 I_V^\prime \ \cos^2 2 \theta^m_0
\nonumber \\
& = & \left( \cos I_V^\prime \ \sin 2\theta^m_0 \ \sin \phi
+   \sin I_V^\prime \ \cos 2 \theta^m_0 \right)^2.
\label{emushift}
\eea

An interesting feature of the obtained results  is that the probabilities for symmetric
transitions:  the flavor-to-flavor and  mass-to-mass ones can be written as a square
of the sum of two terms  proportional to  $\cos I_V$ and $\sin I_V$.

\subsection{Adiabatic  perturbation theory in Magnus expansion}

Let us again consider symmetric density profile.
As the propagation basis,  we take the basis of
the eigenstates of instantaneous Hamiltonian,
$\nu^m \equiv (\nu_{1m}, \nu_{2m})$,:
$$
\nu_f = U(\theta^m(x)) \nu^m~.
$$
Here $\theta^m(x)$ is the instantaneous mixing angle in matter (\ref{mangle}).
The Hamiltonian for the eigenstates
equals $H(x) = H_0 + \Upsilon_\theta (x)$,  where
\be
H_0(x) =\left(
\begin{tabular}{cc}
0 & 0 \\
0 & $\Delta^m (x)$
\end{tabular}
\right) \ , \ \
\Upsilon_\theta (x) = \dot{\theta}^m(x) \left(
\begin{tabular}{cc}
0 & $-i$ \\
$i$ &  0
\end{tabular}
\right),
 \ee
and
\be
\dot{\theta}^m(x) \equiv {d \theta^m(x) \over dx}= {\sin 2
\theta^m(x) \over 2 \Delta^m(x) } ~{d V(x) \over dx}.
\label{thetadot}
\ee
In what follows we will use $H_0$ and  $\Upsilon_\theta (x)$ as
the self-commuting and perturbation parts correspondingly.
Notice that the self-commuting part is the same as before, but
the perturbation is different since the basis of states differs
from the one  we used before.
Now  $\Upsilon_\theta (x)$ is a complex and non-symmetric matrix with
respect to the middle of trajectory.
Straightforward calculations give according to (\ref{cc1122}) or (\ref{c12-mid})
\be
{\bar C}_1 =
I_\theta \left(
\begin{tabular}{cc}
0 & 1 \\
1 &  0
\end{tabular}
\right), ~~~~~
{\bar C}_2 =
I_{\theta\theta}  \left(
\begin{tabular}{cc}
1 & 0 \\
0 & -1
\end{tabular}
 \right),
\label{c1c2}
\ee
where
\bea
I_\theta &=&- 2 \int_{\bar x}^{x_f}  \dot{\theta}^m(x) \ \sin
\phi_{{\bar x} \to x} dx  =
\nonumber \\
&=& 2 \int_{\bar x}^{x_f} \! \! [\theta^m(x)-\theta_s^m] \
\Delta^m (x) \ \cos \phi_{{\bar x} \to x} \ dx ,
 \eea
 \bea
I_{\theta \theta} &=& -\int_{x_0}^{x_f}dx \int_{x_0}^x  {\dot
\theta}^m(x) {\dot \theta}^m(y) \sin \phi_{y \to x} dy = \nonumber
\\
 &=&
 4 \int_{\bar{x}}^{x_f}dx \int_{\bar{x}}^x  {\dot
\theta}^m(x) {\dot \theta}^m(y) \sin \phi_{\bar{x} \to y} \cos
\phi_{\bar{x} \to x} dy .
 \eea
Here $\theta_s^m = \theta^m (x_0) = \theta^m (x_f)$ is the mixing angle
at the surface of the Earth. Taking into account (\ref{thetadot})
one sees that $I_{\theta \theta}$ has the same structure as the
integral in $Y(x)$ (\ref{yyy}).

Neglecting the second order term $\propto I_{\theta \theta}$,
we have $\bar{C} = \bar{C}_1$ which coincides, according to (\ref{c1c2}),
with total $\bar{C}$ in eq. (\ref{barc-11}) up to the  change $I_V^{\prime} \rightarrow
I_\theta$.  Therefore the adiabatic
probabilities equal to those in the previous
subsection with the substitutions $I_V^{\prime} \rightarrow I_\theta$, $\theta^m_0
\rightarrow \theta^m_s $ :
\bea
P_{\nu_2 \to \nu_e} & = &
\cos^2 I_\theta \ \sin^2 \theta \ +
 \cos^2 I_\theta \ \sin 2\theta^m_s  \ \sin 2 (\theta^m_s-\theta) \ \sin^2 \phi \ +
 \nonumber \\
 & + &
  {1 \over 2 } \sin 2 I_\theta \ \sin 2 (2\theta^m_s-\theta) \ \sin \phi \
+ \sin^2 I_\theta \ \cos^2 (2 \theta^m_s - \theta),
\label{mmee}
\eea
\bea
P_{\nu_2 \to \nu_1} &=&
\cos^2 I_\theta \ \sin^2 2 (\theta^m_s -\theta) \ \sin^2 \phi  +
\ {1 \over 2 } \sin 2 I_\theta \ \sin 4 (\theta^m_s -\theta) \ \sin \phi \
\nonumber \\
& + &  \sin^2 I_\theta \ \cos^2 2( \theta^m_s - \theta) \ ,
\label{mmmm}
\eea
\bea \label{eeee}
P_{\nu_e \to \nu_\alpha} & = & \cos^2 I_\theta
\ \sin^2 2\theta^m_s   \  \sin^2 \phi + \ {1 \over 2 } \sin 2
I_\theta  \ \sin 4\theta^m_s \ \sin \phi  +
\nonumber \\
& + & \ \sin^2 I_\theta \ \cos^2 2 \theta^m_s.
\eea
Notice that $I_\theta  \approx I_V^\prime $,  when $\theta^m -\theta^m_s
\ll 1$ .

Let us take into account the second order of the Magnus expansion.
Now $\bar{C}$ contains the term proportional to the
diagonal matrix. Apparently, $\bar{C}$ has the same form as in
(\ref{upscc}) with the substitutions $Z \rightarrow I_{\theta}$ and
$Y \rightarrow I_{\theta \theta}$.
So,  using the results (\ref{smatxyz}), (\ref{c1c2}), (\ref{av-rep}) and
(\ref{s-origin}), we find
the $S$-matrix in the basis of eigenstates of the Hamiltonian,
\be
S_m =
\left(
\begin{array}{cc}
\cos I_t - i \frac{I_{\theta \theta}}{I_t} \sin I_t  &
 -i \frac{I_{\theta}}{I_t}  \sin I_t e^{-i\phi}  \\
 -i \frac{I_{\theta}}{I_t}  \sin I_t e^{-i\phi}  &
\left[ \cos I_t + i \frac{I_{\theta \theta}}{I_t} \sin I_t \right] e^{-2i\phi}
\end{array}
\right).
\label{smat-eig}
\ee
Here
\be
I_t \equiv \sqrt{I_{\theta}^2 + I_{\theta \theta}^2},
\ee
and the adiabatic phase $\phi$ is defined in (\ref{phasesh}).
The $S$-matrix for the flavor-to-flavor transitions is then given by
\begin{equation}
S_{flavor - flavor} = U(\theta_s^m) \cdot S_m \cdot U^{\dagger}(\theta_s^m).
\end{equation}
For the probability of $\nu_e \to \nu_\alpha$ oscillations,
$P_{\nu_e \to \nu_\alpha} = |(S_{flavor - flavor})_{\alpha e}|^2$,
we obtain explicitly
\be
 P_{\nu_e \to \nu_\alpha} =
        \left[ \sin 2 \theta^m_s \  \cos I_t \ \sin \phi \ +
  \ {\sin I_t \over I_t}
        (I_\theta \ \cos 2\theta^m_s - I_{\theta \theta} \ \sin 2 \theta^m_s \ \cos \phi )
\right]^2 .
\label{2ord}
\ee
The $S$-matrix for the mass-to-flavor transitions equals
\begin{equation}
S_{mass - flavor} = U(\theta_s^m) \cdot S_m \cdot U^{\dagger}(\theta_s^m - \theta).
\end{equation}
In particular, the $\nu_2 \to \nu_e$ - probability can be calculated as
$P_{\nu_2 \to \nu_e} = |(S_{mass - flavor})_{e2}|^2$; and explicitly we obtain
\bea
 P_{\nu_2 \to \nu_e} &  =  &
        \left[ \sin (2 \theta^m_s - \theta) \  \cos I_t \ \sin \phi \ +
  \ {\sin I_t \over I_t}
        \left(I_\theta \ \cos (2\theta^m_s - \theta)
- I_{\theta \theta} \ \sin (2\theta^m_s - \theta) \ \cos \phi \right)
\right]^2
\nonumber \\
& + & \sin^2 \theta  \left[\cos I_t \ \cos \phi \ +
 \ {\sin I_t \over I_t} I_{\theta \theta}\sin \phi \right]^2.
\label{2ordexp}
\eea

Notice that the adiabatic perturbation theory is essentially a series in
\be
\frac{\dot{\theta}_m}{\Delta^m} =
\dot{V} \frac{\sin 2\theta \frac{\Delta m^2}{2E}}{2(\Delta^m)^3},
\label{adpar}
\ee
i.e., in  gradient of the potential rather that in $2VE /\Delta m^2$.
Therefore this theory is applied also for
$2VE /\Delta m^2 > 1$. The largest value of the parameter
(\ref{adpar}), at least for small vacuum mixing,
is achieved in the MSW-resonance, where
\be
\frac{\dot{\theta}_m}{\Delta^m} =
\frac{1}{2\pi}\frac{\dot{V}}{V}\frac{l_\nu}{2 \sin2\theta \tan 2\theta}
= \frac{1}{2\pi}\frac{l_\nu^{res}}{\Delta r_R}.
\label{adpar-r}
\ee
Here $l_\nu \equiv 4\pi E/\Delta m^2$  and
$l_\nu^{res} \equiv l_\nu/\sin 2\theta$
are the oscillation lengths in vacuum and in matter with the  resonance density,
$\Delta r_R \equiv   2 \tan 2\theta (V/\dot{V})$
is the spatial  width
of the resonance layer. So, the approximation is not expected to  work
well in resonance for small mixing.

\section{Accuracy of semi-analytic approximations}

\begin{figure}
\includegraphics[height=90mm]{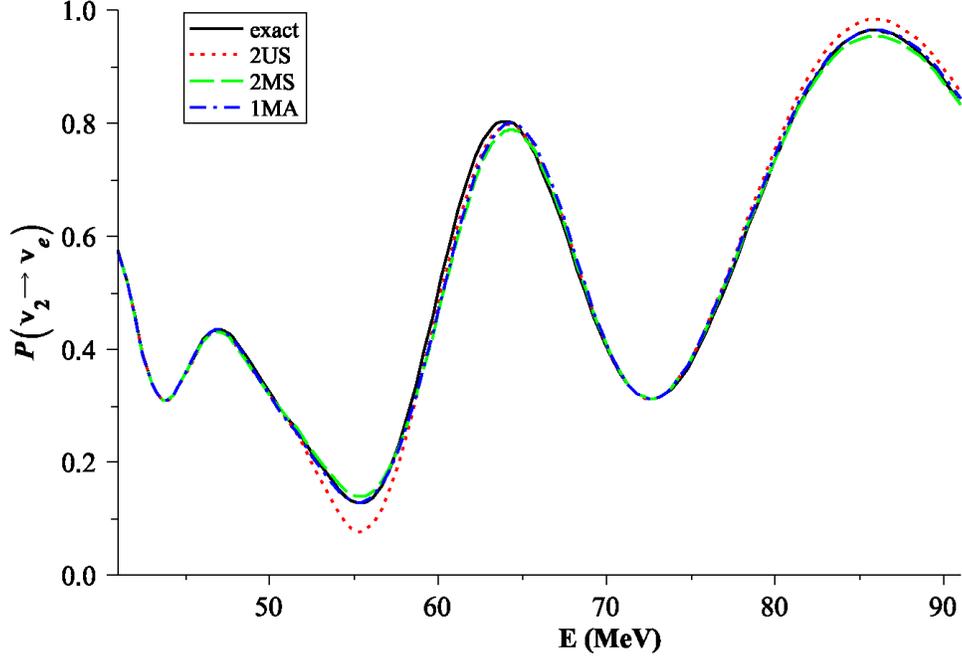}
    \caption{\label{f12}%
The  $\nu_2 \rightarrow \nu_e$ oscillation probabilities driven by
the oscillation parameters: $\Delta m^2 = 7 \cdot 10^{-5}$ eV$^2$
and $\sin^2 \theta_{12} = 1/3$ as functions of the neutrino
energy. The lines correspond to the exact numerical computations
(solid), the second order of usual (non-unitary) perturbation
theory (dot-dashed); the 2nd order Magnus expansion with shifted
potential (dashed), the first order adiabatic Magnus expansion
(dotted).}
\end{figure}
\begin{figure}
\includegraphics[height=90mm]{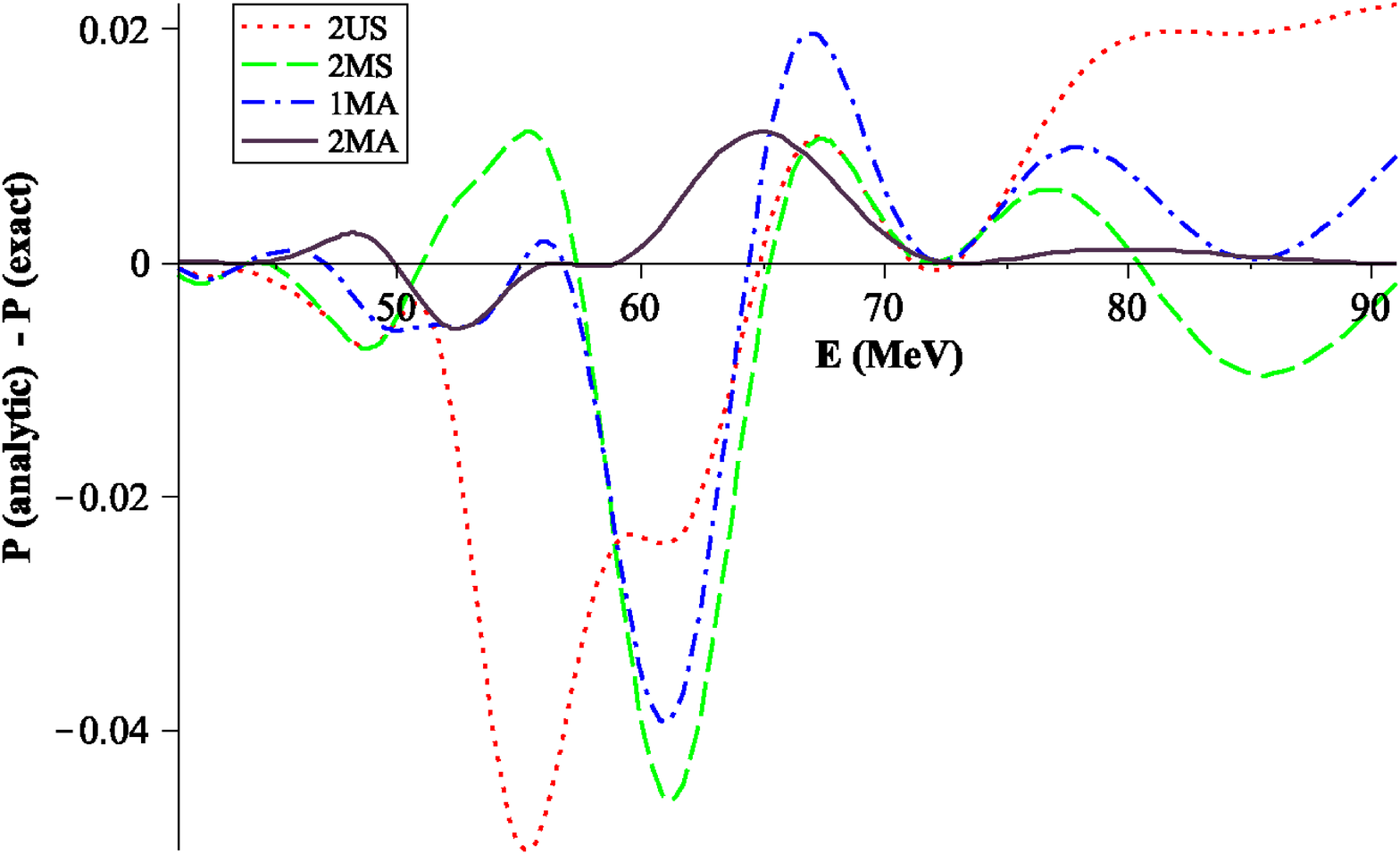}
    \caption{\label{f12d}%
The deviation of the approximate value of the $\nu_2 \rightarrow
\nu_e$ probability from the exact value as a function of neutrino
energy. The lines correspond to the second order of usual
(non-unitary) perturbation theory (dot-dashed); the second order
Magnus expansion with shifted potential (dashed); the first order
adiabatic Magnus expansion (dotted); the second order
adiabatic Magnus expansion (solid).
}
\end{figure}

To illustrate an accuracy of the obtained semi-analytical results we
consider neutrino oscillations along the trajectory which crosses
the center of the Earth (the central trajectory).
We take the 5-layer approximation for the Earth density profile \cite{LM}.
We compute  $P_{exact}$ using
exact numerical method,  and
$P_{analytic}$ -
the approximate probabilities, using  different semi-analytic
formulas obtained in this paper.
The Table I lists approximations we use to produce
the figures with indication of abbreviations and references to
the corresponding formulas in the text.

\begin{center}
\begin{table}[h]
\caption{\label{osc-tab}
Approximations}
\centering
\begin{tabular*}{\columnwidth}{@{\extracolsep{\fill}}@{}lll@{}}
\hline
Notation & Approximation &  Equation\\
\hline
1MA  & first order Magnus adiabatic expansion & (\ref{mmee})\\
2US  & second order usual expansion with shifted  potential &
(\ref{mf-shift1})\\
2MS  & second order Magnus expansion with shifted potential & (\ref{mf-shift}) \\
2MA &  second order Magnus adiabatic assumption & (\ref{2ordexp})\\
\hline
\end{tabular*}
\end{table}
\end{center}

In fig.\ref{f12} we show the  probabilities of $\nu_2 \rightarrow \nu_e$
oscillations driven by the parameters $\Delta m^2 = 7 \cdot
10^{-5}$ eV$^2$ and $\sin^2 \theta_{12} = 1/3$. Fig. \ref{f12d}
presents the differences of  the semi-analytic and exact results,
\be
\Delta P \equiv P_{analytic} - P_{exact},
\ee
as functions of the neutrino energy.
The solid  line in Fig. \ref{f12} shows $P_{exact}$.
Apparently, the probabilities and the differences of probabilities
increase with energy; the probabilities become of the order 1 in the resonance
region $E\sim 100$ MeV.

Let us discuss the  quality of different approximations.

\begin{itemize}

\item

The dot-dashed  lines  show $P_{analytic}$ (fig.~\ref{f12})  and
$\Delta P$ (fig.~\ref{f12d}) computed in the second order of the
usual perturbation theory in   (practically in  $I_V^\prime$) with a
shifted potential, $2US$. The probability is
given by  an expansion of the expression (\ref{mf-shift}) in powers of
$I_V^\prime$:
\bea
 P_{\nu_2 \to
\nu_e} = \sin^2 \theta
 + \sin 2\theta^m_0 \sin 2(\theta^m_0-\theta)\sin^2 \phi +
 I_V^\prime  \sin 2 (2\theta^m_0-\theta) \sin \phi
+
\nonumber\\
+ (I_V^\prime)^2  [\cos^2 (2 \theta^m_0- \theta) - \sin^2 \theta -
\sin 2\theta^m_0 \sin 2(\theta^m_0-\theta)\sin^2 \phi].
\label{mf-shift1}
 \eea
This probability coincides with our result  in
\cite{Ioannisian:2004vv}. We use the average value of potential,
$V_0$,  that corresponds to the electron density $n_e = 1.92 \ N_A \
mol/cm^3$, where $N_A$ is the Avogadro number.
For the central  trajectory
the probability (\ref{mf-shift1}),
satisfies inequality $P \leq 1$.
However, for some  other trajectories, e.g., with the nadir angle
$\Theta \sim 10^{\circ}$, the unitarity is violated.

\item
In the first order of usual perturbation theory, the probability is
given by eq. (\ref{mf-shift1}) without last term (the line is not shown in the figure).
It becomes $P > 1$ for the central trajectory
in the region (80 - 90) MeV reflecting the violation of unitarity.

\item
The dashed  line in fig.~\ref{f12}  shows  the probability (\ref{mf-shift})
computed in the second order Magnus expansion, $2MS$,
with  $\Delta V$ ($I_V^\prime$)  as the perturbation.
The unitarity is  restored and  $P \leq 1$ for all  energies
and for all the trajectories.
The difference of probabilities $\Delta P$ is shown in fig. \ref{f12d}.
At high energies  (large $I_V^\prime$) this probability gives
substantially better approximation than the non-unitary one:
the deviation is below 5 \%. At low energies, $E < 45$ MeV, (small $I_V^\prime$)
both approximations have similar accuracy.
As follows from the figure the deviation $\Delta P$ at high energies becomes even
smaller: below  2\% in the range 80 - 100 MeV.

\item
$P_{analytic}$ and $\Delta P$ calculated in the adiabatic
perturbation theory in the first order of the Magnus expansion
$1MA$ (\ref{mmee}) are shown by the dotted lines. According to  the figures a
quality of the first order adiabatic approximation with restored unitarity
is similar to that of the second order in the
$\Delta V-$ perturbation theory (the previous case).
This means that the adiabatic perturbation theory is more relevant
in combination with the Magnus expansion than the usual perturbation theory
(practically in $VE/\Delta m^2$). The adiabatic perturbation theory gives better
re-summation of the series.
The comparable qualities of these approximations
are related also to the fact that in both cases
we have taken the same values of the electron density:
the surface density in the adiabatic case
and the average density in the $\Delta V-$ perturbation,
so that  $ \theta_s =  \theta_0 $. Furthermore, as we have mentioned in the sec. IIIC.
the true expansion parameter is  $\dot{\theta}_m \propto \dot{V}$,
and the perturbation theory works also for $VE/\Delta m^2 \geq 1$.

\item
The solid line in fig. \ref{f12d} shows the difference of
probabilities for  $P_{analytic}$
computed in the second order of the adiabatic Magnus expansion (\ref{2ordexp})
$2MA$. The second order expansion further improves approximation
for all energies and especially in the range (50 -65) MeV and at
$E > 70$ MeV.
In fact, the approximation works well in whole energy
range: below the resonance, in the resonance and above. The adiabaticity is
well satisfied due to large value of the vacuum mixing angle.
To illustrate this in fig.~\ref{he}
we show the probabilities computed in different
approximations at high energies - above the resonance.

\begin{figure}
\includegraphics[height=90mm]{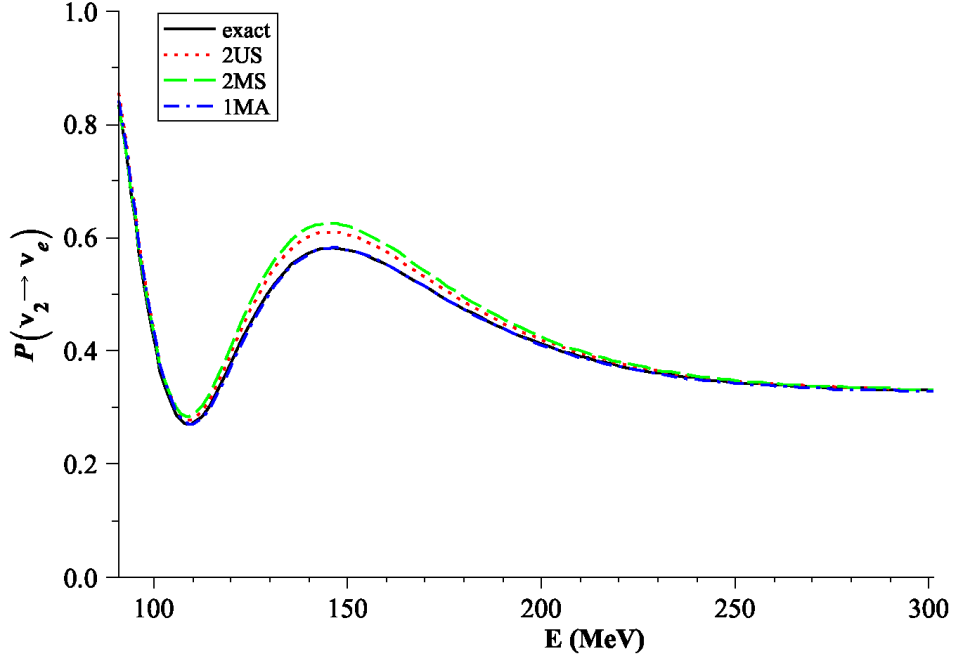}
    \caption{\label{he}%
The same as in fig. \ref{he} for high energy range.}
\end{figure}



\end{itemize}


\begin{figure}
\begin{center}
\includegraphics[height=90mm]{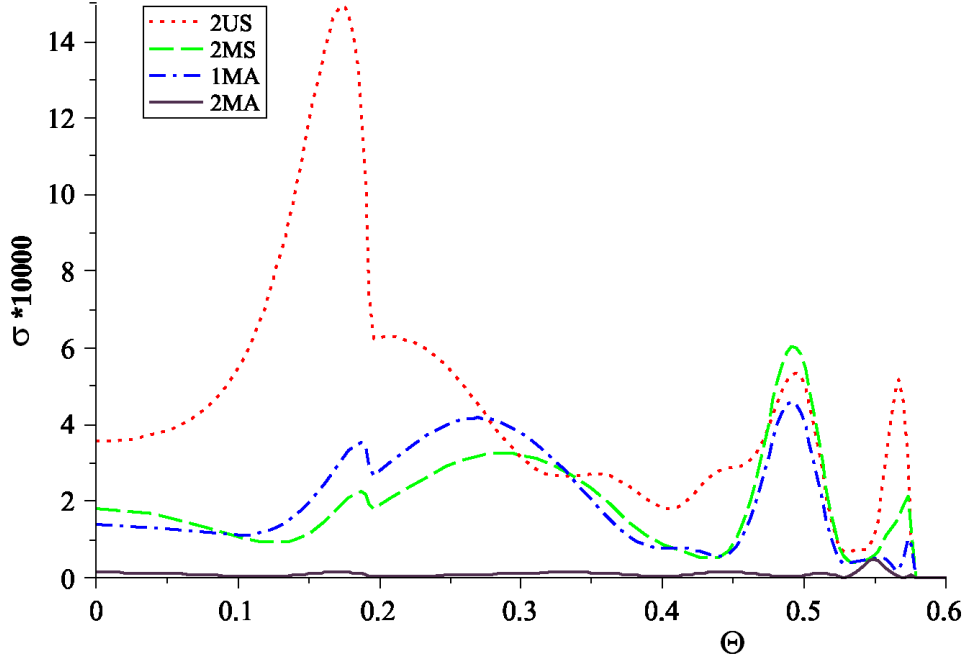}
    \caption{\label{f12an}%
Dependence of the  energy integrated errors of various approximations
on the nadir angle (in radians) of neutrino trajectory.
The errors (in units $10^{-4}$) are computed for
the  $\nu_2 \rightarrow \nu_e$ oscillation channel with
parameters $\Delta m^2 = 7 \cdot 10^{-5}$ eV$^2$ and $\sin^2 \theta_{12} = 1/3$.
The lines correspond to different approximations as in fig. \ref{f12d}.
}
\end{center}
\end{figure}

Similar picture appears for other neutrino trajectories.
In fig. \ref{f12an} we show dependence of  the integral error of the approximations
defined as
 \be
\sigma \equiv {1 \over E_{max}-E_{min}} \int_{E_{min}}^{E_{max}}
(P_{analytic}  - P_{exact})^2 d E
 \ee
on  the nadir angle $\Theta$ (in radians). We show the range of the angles which
corresponds  to the core crossing trajectories;
$\Theta = 0$  determines the central trajectory considered above.
We take $E_{min}=40$ MeV and $E_{max} = 90$ MeV. As follows from the figure,
the Magnus expansion gives much better approximation than the usual (non-unitary)
perturbation theory.
Again, the accuracy of the first order adiabatic Magnus expansion $1MA$ and
the second order Magnus expansion in $\Delta V$,  $2MS$, are comparable.
The second order adiabatic Magnus expansion, $2MA$, gives much better approximation
for all the energies.

According to fig.~\ref{f12an}, the errors become very small for
$\Theta \rightarrow 0.58$ which corresponds to the only-mantle
crossing trajectories.  This means that the proposed approximations have
even higher accuracy for neutrinos propagating only in the mantle.\\

\begin{figure}
  \includegraphics[height=90mm]{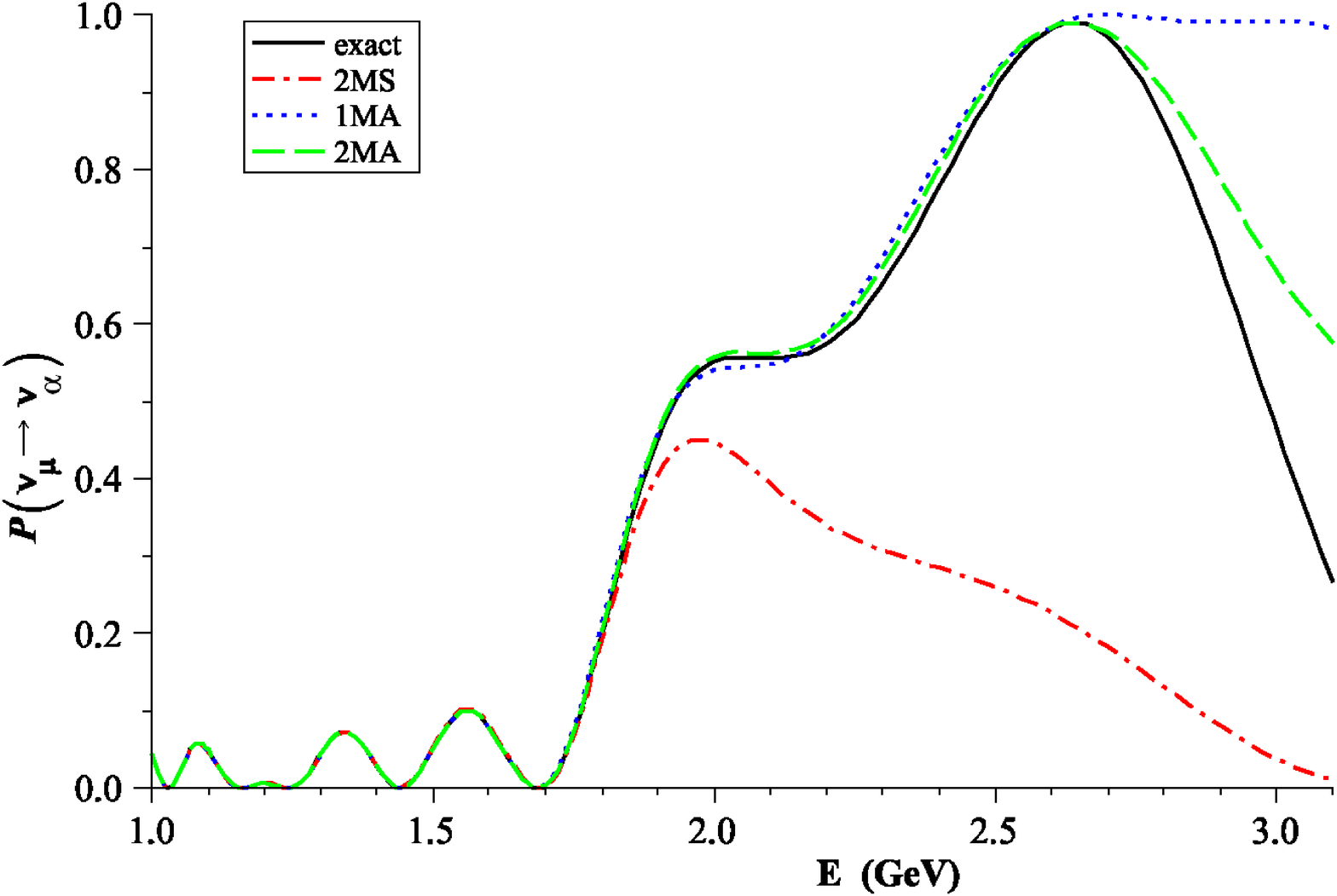}
    \caption{\label{f13}%
The probabilities of the $\nu_e \rightarrow \nu_\alpha$ transition
as functions of neutrino energies computed in various
approximations. The lines correspond to the exact numerical
calculations (solid), the 2nd order of usual non-unitary expansion
(dot-dashed), the first order adiabatic Magnus expansion (dotted)
and the second order adiabatic Magnus expansion (dashed). The
values of oscillation parameters are $\Delta m^2 = 2 \cdot
10^{-3}$ eV$^2$ and $\sin^2 \theta_{13} = 0.01$. }
\end{figure}

In fig. \ref{f13} we compare the $\nu_e \rightarrow \nu_\alpha$
probabilities  due to $\Delta m^2 = 2 \cdot
10^{-3}$ eV$^2$ and $\sin^2 \theta  = 10^{-2}$. The solid line is
the result of exact computations.
Comments on the  accuracy of different approximations follow.

\begin{itemize}

\item
The dot-dashed line is a result of the
second order of the usual non-unitary $\Delta V-$ perturbation theory. It
corresponds to  expansion of the probability (\ref{emushift}):
\be
 P_{\nu_e \to \nu_\alpha}  =
\sin^2 2\theta^m_0 \ \sin^2 \phi +  I_V^\prime \ \sin
4\theta^m_0 \ \sin \phi  + (I_V^\prime)^2  (\cos^2 2\theta^m_0  -
\sin^2 2\theta^m_0  \sin^2 \phi).
 \ee
The approximation work well at $E < 1.7$ GeV, where the probability is small
$P < 0.3$. For higher energies it fails completely.
According to the figure at $E > 3$ GeV this
probability becomes negative indicating a violation of the unitarity.

\item
The green dotted line shows the probability in the first order of
the adiabatic Magnus expansion (\ref{eeee}) $1MA$.
The Magnus expansion allows us to expand the region up to
2.7 GeV, {\it i.e.} practically up to the resonance in the core of the Earth.

\item
The dash-dotted line represents the probability in the second order of the
adiabatic Magnus expansion (\ref{2ord}), $2MA$.
It has even better accuracy: For $E
= 2.3$ GeV  we obtain  $\Delta P \sim 0.05 $ for the first order  and
$\Delta P \sim 0.02 $ - for the second one. The approximation
becomes invalid for $E > 2.8$ GeV because the adiabaticity is
broken in the resonance.

\end{itemize}

Let us underline that the Magnus expansion allows one
to extend the application of approximation to the region where the
probabilities are large.
The semianalytic result does not work in the resonance region.
It gives good approximation above 8 GeV, that is,  above
the resonance in the mantle.

\begin{figure}
  \includegraphics[height=90mm]{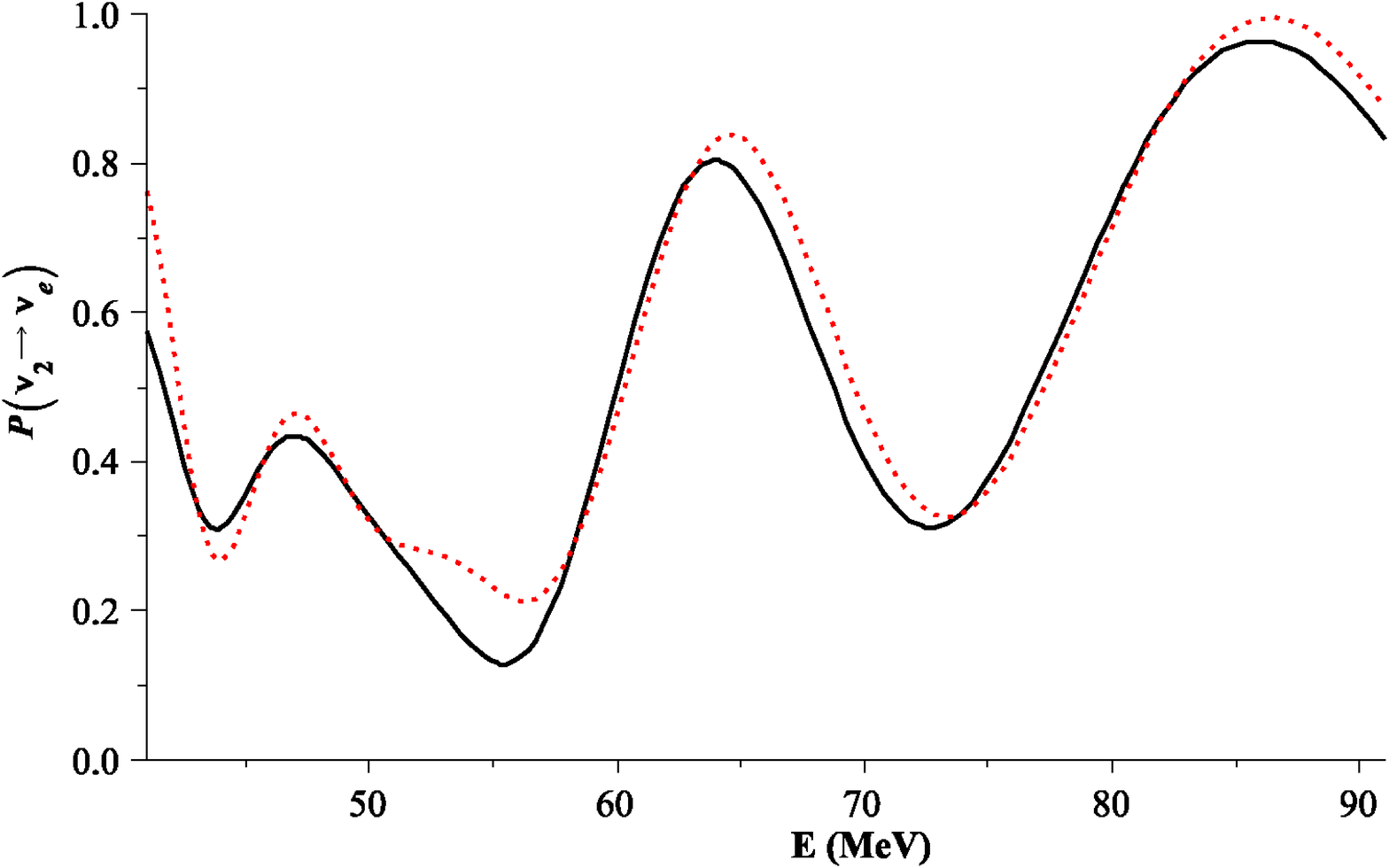}
    \caption{\label{2layer}%
The probabilities of the $\nu_2 \rightarrow \nu_e$ transition as
functions of neutrino energies computed for the exact (5 layers)
Earth matter density profile (solid line),  and for the two-layer
approximation of the profile (dotted line).}
\end{figure}

Let us compare an accuracy of our semi-analytic results with the
exact results of calculations for the widely used two-layer
density approximation of the Earth profile \cite{3lay}. In this
approximation the densities of the mantle and the core of the
Earth are taken to be constant  and equal to the mean densities in
the mantle and the core along a given neutrino trajectory.
Fig.~\ref{2layer} shows the $(\nu_2 \rightarrow \nu_e)$
probabilities for the central trajectory for the exact (5 layers)
density profile (solid line) and the two-layer density
approximation (dotted line). At high energies, $E = 60 - 70$ MeV,
the accuracy of approximation is about 4 - 5 \%. The accuracy
becomes worser with a decrease of energy: at $E \sim  45$ MeV  and
below, it is about (20 - 30) \%. Partly the loss of accuracy is
related to the fact that at low energies the propagation becomes
more adiabatic and therefore the result of propagation in the
mantle is determined by the density at the surface of  mantle,
rather than the average density.

Comparing  fig. \ref{2layer} and fig. \ref{f12} we conclude that
the semianalytic results based on the Magnus expansion give
better approximation outside the resonance regions than
the exact results obtained for the two-layer model of the Earth density profile.\\

Let us finally comment on embedding of our $2\nu-$results in  the  complete
$3\nu$-mixing framework.
In certain limits relevant for applications the dynamics of
$3\nu-$system is reduced to the dynamics of $2\nu-$system.
These include  the limits of
low energies (substantially below the 1-3 resonance energy), and
high energies (substantially larger than the 1-2 resonance energy).

Let us consider the low energy case, $E < 100$ MeV.
At low energies one can neglect the matter effect on the 1-3 mixing,
and furthermore, the  oscillations related to the third mass eigenstate
(separated by the atmospheric $\Delta m^2$) are averaged out.
This state essentially decouples from the dynamics
and evolves independently.
In this case it is  straightforward to show that, e.g., the $3\nu-$probability of
the $\nu_2 \rightarrow \nu_e$ transition,
$P_{\nu_2 \rightarrow \nu_e}^{(3\nu)}$,  is given by
\be
P_{\nu_2 \rightarrow \nu_e}^{(3\nu)} = \cos^2 \theta_{13} P_{\nu_2 \rightarrow \nu_e}
(\theta_{12}, \Delta m^2_{21}, V \cos^2 \theta_{13}).
\ee
Here $P_{\nu_2 \rightarrow \nu_e}$ is the two neutrino probability
derived in this paper, (see  eqs. (\ref{2eresum}, \ref{mf-shift},
\ref{mmee})) which should be computed using the  reduced value of the potential: $V \cos^2
\theta_{13}$.

\section{Conclusions}

We have developed new formalism of computations of the oscillation probabilities in
matter with varying density. It is based on the Magnus expansion and has a virtue
to be unitary in each order of the expansion.
The formalism can be adjusted to a specific physical situation  by
choosing a neutrino evolution basis and a split of the Hamiltonian into
the self-commuting and non-commuting parts.
The latter can be used as a perturbation.
Using the Magnus expansion one can develop different perturbation theories,
and in particular, the improved adiabatic perturbation theory.
The evolution due to self-commuting part can be accounted for
in a way which  is equivalent to a transition to the ``interaction representation'' in
quantum mechanics. \\

We have obtained  the semi-analytical formulas
for various oscillation probabilities in the second order of the Magnus expansion.
The Magnus expansion (apart from being unitary) leads also to better convergence
of series. We show that the Magnus expansion corresponds to certain re-summation
of contributions in the  usual perturbation theory, and it is this re-summation that
leads to restoration of unitarity. The developed unitary formalism gives
new insight into the previously obtained results and their limitations.\\

Using several explicit examples  we show that the restoration of unitarity gives
better approximation to the results of exact numerical calculations, especially in the
region where the transition probabilities are large.
We find that the best approximation (among the considered examples)
is provided by the adiabatic Magnus expansion. \\

The results  in sec. II and III have a general character
valid for wide class of potentials not necessarily related to the Earth density profile.
Using the proposed method one can develop  other perturbation
approaches adjusting
to particular physical conditions
the  evolution basis and split of
the Hamiltonian.  For instance, at high energies one can use
the matter part of the Hamiltonian as the self-commuting part: $H_0 = diag(V, 0)$,
and the vacuum (kinetic) part  as a perturbation. This
theory will give  good approximation at  high energies,
where $V > \Delta m^2/2E$. \\

We have illustrated our results computing the oscillation probabilities
for neutrinos crossing the core of the Earth (actually most of the figures
are produced for the central trajectory). We find that for
the solar oscillation parameters,  $\Delta m^2_{21}$ and
$\theta_{12}$,   the second order of the Magnus
adiabatic expansion gives a very good precision ($< 1\% $) for all energies.
For the mantle-only trajectories the precision is even higher.
For the atmospheric parameters  $\Delta m^2_{31}$ and small 1-3 mixing
the approximation works well
($< 3\% $ accuracy for $\sin^2 \theta_{13} = 0.01$) below
($E < 2.7$ GeV) and above ($E > 8$ GeV)
the resonance region. In the region $(2.7 - 8)$ GeV the MSW-resonances
in the core and in the mantle as well as the parametric resonances
take place and the Magnus adiabatic approximation fails
since the adiabaticity is broken. In this region one should
use some other approach. For the mantle-only crossing trajectory
the approximation fails in the region $(5 - 8)$ GeV for $\sin^2 \theta_{13} = 0.01$.\\

The results obtained here can be used for description of
propagation of the solar and supernova neutrinos inside the Earth.
They also can be used to describe the flavor oscillations
of the atmospheric and accelerator  neutrinos.
For solar neutrinos, $E < 18$ MeV, the transition probability is small,
so that already usual perturbation theory gives very good approximation.
The Magnus expansion adds little,  as far as accuracy is concerned.
For the galactic supernova the detectable tail of the energy spectrum
extends up to 50 - 70 MeV (depending on a distance to supernova
and a size of  detector).
The range of energies $E > 40$ MeV, where the Earth matter effect is
enhanced, is of special interest both for measurements of the neutrino
parameters and for physics of gravitational collapse and
mechanism of star explosion. It is this range where the Magnus
expansion gives substantial improvement of accuracy.
For the atmospheric neutrinos, the Magnus adiabatic approximation can be used to
describe oscillations driven by the 1-2 mass split and 1-2 mixing for
all neutrino energies and all trajectories. It is especially  relevant
for low energies: the sub-GeV events as well as events below 100 MeV.
The results can be applied for oscillations induced by the
1-3 mass split and 1-3 mixing outside the resonance regions.
They can be used for long baseline experiments with neutrino
energies below 3 GeV (thus covering the range of proposed superbeams)
and for high energy beams from neutrino
factories ($E > 8$ GeV). The results can be applied for
neutrinos of cosmic origin.

\section{Appendix}

The functionals $C_k[H]$ can be derived in the following way.
The standard  expansion of the chronological product
\bea
 T \ e^{-i\int_{x_0}^{x_f} H(x) \ dx }
 &=&1  - i \int_{x_0}^{x_f} \!\!\!\!\!dx \ H(x) + (-i)^2\int_{x_0}^{x_f}
\!\!\!\!\!dx \int_{x_0}^x \!\!\!\!\!dy \ H(x) H(y) \nonumber \\
 && + (-i)^3
\int_{x_0}^{x_f} \!\!\!\!\!dx \int_{x_0}^x \!\!\!\!\!dy
\int_{x_0}^y \!\!\!\!\!dz \ H(x) H(y) H(z) + \cdots
 \label{t-prod}
 \eea
can be rewritten in terms of the commutators of the Hamiltonian using the
following identities
\bea
\label{iden1}
 &&\int_{x_0}^{x_f} \!dx \int_{x_0}^x \!dy \, H(x) H(y)
\equiv  {1 \over 2 }\int_{x_0}^{x_f} \!dx
\int_{x_0}^x \!\!\!\!\!dy \ [ \ H(x), \  H(y) \ ] + {1 \over 2
}\left(\int_{x_0}^{x_f} \!\!\!\!\!dx \
H(x)\right)^2  \ ,  \\
&& \int_{x_0}^{x_f} \!\!\!\!\!dx \int_{x_0}^x \!\!\!\!\!dy
\int_{x_0}^y \!\!\!\!\!dz \ H(x) H(y) H(z) = \nonumber
\\
&& \equiv {1 \over 6 }\int_{x_0}^{x_f} \!\!\!\!\!dx \int_{x_0}^x
\!\!\!\!\!dy \int_{x_0}^y \!\!\!\!\!dz \  \{  \left[ \ H(x),\ [ \
H(y), \ H(z) \ ] \right] + \left[[ \ H(x), \ H(y)\ ], \
H(z) \right] \} \nonumber \\
&& \ \ \ + {1 \over 2 } \left\{ \int_{x_0}^{x_f} \!\!\!\!\!dx \ H(x)
~ \int_{x_0}^{x_f} \!\!\!\!\!dx \int_{x_0}^x \!\!\!\!\!dy \ [
\ H(x), \  H(y) \ ] + \int_{x_0}^{x_f} \!\!\!\!\!dx \int_{x_0}^x
\!\!\!\!\!dy \ [ \ H(x), \  H(y) \ ] ~ \int_{x_0}^{x_f}
\!\!\!\!\!dx \ H(x) \right\} \nonumber
\\
&& \ \ \  +  {1 \over 6 }\left(\int_{x_0}^{x_f} \!\!\!\!\!dx \
H(x)\right)^3 \ ,
\label{iden2}
\end{eqnarray}
etc..
These identities follow from an extension of all the integrations over whole
range from $x_0$ to $x_f$. For instance,  eq. (\ref{iden1})
can be derived taking into account that in the double integral over $x$ and $y$
$$
I(y = x_0 \div x ) + I (y = x \div x_f) = \left[\int_{x_0}^{x_f} dx  H(x)\right]^2,
$$
and on the other hand
$$I (y = x \div x_f) =  I(y = x_0 \div x) -
\int_{x_0}^{x_f} \int_{x_0}^x dx dy [H(x), H(y)].
$$
Inserting (\ref{iden1}) and (\ref{iden2}) into (\ref{t-prod})
we obtain
\be
 S= 1  - i \int_{x_0}^{x_f} \!dx \ H(x) +
(-i)^2 {1 \over 2} \int_{x_0}^{x_f}\!dx \int_{x_0}^x \!dy \ [ H(x), H(y)]
+ (-i)^2 {1 \over 2} \left( \int_{x_0}^{x_f} \!dx \ H(x) \right)^2
+ \cdots ,
\label{expcom}
\ee
where we have written explicitly the commutators up to the third order.
On the other hand expanding (\ref{Crepres1})
we have
\be
S = 1 - i (C_1 + C_2 + C_3) + \frac{(-i)^2}{2}( C_1^2 + 2 C_1 C_2)
+ \frac{(-i)^3}{6} C_1^3 + \cdots
\label{exp-c}
\ee
Comparing (\ref{expcom}) and (\ref{exp-c}) we
obtain immediately the results (\ref{c111}, \ref{c222}, \ref{c333}).

\vspace{1cm}

The work of A.I. was supported by the NFSAT grant No.
ARP2-3234-Ye-04.

\end{document}